\documentclass[11pt,tightenlines,eqsecnum,floats,aps,amsmath,amssymb,nofootinbib,prd,shownopacs,floatfix]{revtex4}

\usepackage{amsmath,amssymb,amsfonts,dcolumn,color,graphicx,graphics,latexsym}
\usepackage[mathscr]{eucal}
\usepackage[latin1]{inputenc}
\usepackage{enumerate}
\newcommand{\be}{\begin{equation}}
\newcommand{\ee}{\end{equation}}
\newcommand{\ba}{\begin{eqnarray}}
\newcommand{\ea}{\end{eqnarray}}

\newcommand{\lp}{l_{\mathrm{Pl}}}
\newcommand{\f}{\frac}

\newcommand{\bmult}{\nopagebreak[3]\begin{multline}}
\newcommand{\emult}{\end{multline}}

\def\d{{\rm d}}
\def\p{\partial}

\begin{document}

\title{Geodesic completeness and the lack of strong singularities in effective loop quantum Kantowski-Sachs spacetime}
\author{Sahil Saini}
\email{ssaini3@lsu.edu}
\author{Parampreet Singh}
\email{psingh@phys.lsu.edu}
\affiliation{ Department of Physics and Astronomy,\\
Louisiana State University, Baton Rouge, LA 70803, U.S.A.}

\begin{abstract}
Resolution of singularities in the Kantowski-Sachs model due to non-perturbative quantum gravity effects is investigated. Using the 
effective spacetime description for the 
improved dynamics version of loop quantum Kantowski-Sachs spacetimes, we show that even though expansion and shear scalars are universally 
bounded, 
there can exist events where curvature invariants can diverge. However, such events can occur only for very exotic equations of state when 
pressure or derivatives of energy density with respect to triads  become infinite at a finite energy density. In all other cases curvature 
invariants are proved to remain finite for any evolution in finite proper time. We find the novel result that all strong singularities are 
resolved for arbitrary matter. Weak singularities pertaining to above potential curvature divergence events can exist. The effective 
spacetime is found to be geodesically complete for particle and null geodesics in finite time evolution.  Our results add to a growing evidence for generic 
resolution of strong singularities using effective dynamics in loop quantum cosmology by generalizing earlier results on  isotropic and Bianchi-I spacetimes.

\end{abstract}

\maketitle

\section{\bf Introduction}

In general relativity (GR), occurrence of singularities brings forth the underlying limitations of the classical continuum spacetime. 
A singular event in classical theory has many characteristics, generally captured via curvature divergences and break down
of geodesic evolution in a finite proper time. 
However, not all singularities are necessarily the boundaries of classical spacetime. Their strength matters. It is believed that strong 
singularities \cite{Ellis,tipler,krolak}, which cause inevitable complete destruction of arbitrarily strong
in-falling detectors, 
are the true boundaries of the classical continuum spacetime. Strong singular events, such as the big bang or a central singularity inside 
a  black hole, are conjectured to be associated with geodesic incompleteness \cite{tipler,krolak}. 
In contrast, weak singularities can be considered harmless. Even though some curvature components may diverge at such events, 
a sufficiently strong detector survives such a singularity. Geodesics can be extended beyond such singularities in the classical 
spacetime.\footnote{For examples in cosmological spacetimes, see Ref. \cite{jambrina}.} Thus, not all space-like singularities 
are necessarily harmful. A fundamental challenge for  any theory going beyond Einsteinian gravity is whether it can resolve all of the strong 
singularities. 

In the absence of quantum gravitational effects which profoundly modify the structure of the underlying spacetime, singularity resolution has been elusive. In the last decade, applications of loop quantum gravity (LQG) to cosmological spacetimes in loop quantum cosmology (LQC) indicate that 
non-perturbative quantum gravitational modifications play an important role in singularity resolution \cite{as}. Various examples of 
cosmological spacetimes, including  isotropic \cite{aps3,all-iso} and anisotropic models \cite{all-aniso,pswe}, and also hybrid 
Gowdy models \cite{hybrid,hybrid1,gowdy},  have been thoroughly studied at a rigorous quantum level. Extensive numerical simulations of quantum 
cosmological models have been performed \cite{aps3,lsu}, which show that the singularities such as the big bang and big crunch are resolved 
and replaced by a non-singular bounce. Singularity resolution has also been understood in terms of quantum probabilities using consistent 
histories approach \cite{consistent}. 

At the fundamental level, geometry in LQC is discrete resulting in a classical continuum quickly below the Planck curvature scale. 
In fact, the quantum Hamiltonian constraint in LQC on quantum geometry can be very well approximated by the Wheeler-DeWitt Hamiltonian 
constraint on classical continuum as soon as the spacetime curvature becomes less than a percent of the Planck curvature. Interestingly, 
numerical simulations show that an effective quantum continuum description exists which captures the quantum dynamics in LQC at all scales. This effective spacetime description, or effective dynamics, has been used in a plenty of investigations. In relation to the objectives of this manuscript, notable results include the following. Using effective dynamics, expansion and shear scalars have been found to be generically bounded for isotropic and anisotropic spacetimes \cite{cs09,ps09,ps11,gs1,pswe,ps15,ck-bianchi9}. Strong singularities are shown to be absent and effective spacetime is found to be geodesically complete for 
loop quantized isotropic cosmological and Bianchi-I spacetimes for matter with a vanishing anisotropic stress \cite{ps09,ps11,ps15}. However, weak and non-curvature singularities can exist \cite{ps09,ps11,ps15}, of which various examples have been studied in isotropic and Bianchi-I spacetimes in LQC \cite{sudden}.

The goal of this manuscript is to investigate the resolution of strong singularities and geodesic completeness in the Kantowski-Sachs 
model in LQC using the effective spacetime description. We wish to understand under what conditions, for arbitrary non-viscous matter, quantum geometry effects as understood in 
LQC lead to a generic resolution of strong singularities, and whether there exist any weak singularities. Essentially, our aim is to 
generalize the results of geodesic completeness and generic resolution of strong singularities in isotropic and Bianchi-I spacetime in LQC 
to the Kantowski-Sachs model. The Kantowski-Sachs spacetime is an interesting avenue to study for various reasons. In the absence of 
matter, it captures the interior $(r < 2m)$ of the Schwarzschild black hole. In the presence of matter, it is an anisotropic cosmological 
model with a spatial curvature. This spacetime has been loop quantized with different 
regularizations of the Hamiltonian constraint, given by Ashtekar-Bojowald \cite{ab} (see also Refs. \cite{modesto,pullin}), 
Corichi-Singh \cite{cs15}, and Boehmer-Vandersloot \cite{bv} (see also Ref. \cite{chiou}). The first quantization prescription is a  
reminiscent of `fixed area of the loop' procedure in LQC \cite{abl,aps} which is known to have various phenomenological issues due to 
fiducial cell dependence \cite{cs08}. The second and third quantization prescriptions overcome this limitation in their unique ways, and 
give qualitatively different physics of singularity resolution. In this manuscript, we study the Kantowski-Sachs spacetime with 
Boehmer-Vandersloot prescription which is an avatar of the improved dynamics prescription in LQC \cite{aps3}. It should be noted that 
resolution of strong singularities and geodesic completeness for isotropic and Bianchi-I spacetime in LQC have been achieved for this 
particular prescription. Interestingly, a loop quantization of this spacetime, in absence of matter and also in presence of cosmological 
constant, results in a singularity resolution with a pre-bounce spacetime which is a product of two constant curvature spaces and with an 
almost Planckian curvature \cite{jds}. An important result pertinent to our investigations is that the expansion and shear scalars in this 
quantization turn out to be universally bounded \cite{js1}. In the following, our reference to loop quantized Kantowski-Sachs model will 
imply Boehmer-Vandersloot prescription.

Our analysis assumes the validity of the effective dynamics in LQC at all scales. In LQC, effective Hamiltonian
is obtained using a geometrical formulation of quantum theory, and, as noted earlier, turns out to be an excellent approximation for 
isotropic and anisotropic models. This is true at least for states which correspond to  macroscopic spacetimes at late times. A specific example is the 
case of homogeneous and isotropic spatially flat spacetimes where effective Hamiltonian has been derived explicitly using coherent states for the case of the massless scalar \cite{vt}. The resulting effective dynamics has been tested rigourously using numerical simulations \cite{aps3,lsu}, which validate the 
analytical derivation of the effective Hamiltonian for the above family of physical states. In terms of 
the gravitational phase space variables, Ashtekar-Barbero connection components and conjugate triads, the effective Hamiltonian contains 
trigonometric terms of the connections, apart from the triads and the matter variables. These trigonometric or the polymerized terms arise 
from expressing field strength of the connection in terms of holonomies over a minimum physical area in the improved dynamics. The resulting Hamilton's equations from the effective Hamiltonian encode the quantum gravitational repulsiveness and result in 
singularity resolution. In the effective Hamiltonian, there can also be modifications coming from expressing inverse powers 
of triads in terms of Poisson brackets between holonomies and triads. The role of these modifications in singularity resolution is 
generally found to be negligible when compared to the effects originating from the polymerized terms.\footnote{These modifications 
in absence of polymerized terms can also lead to singularity resolution and interesting phenomenology in spatially curved models, see e.g. \cite{st,ms}.} In our analysis, we ignore the latter modifications.\footnote{It should be noted that in models where these terms have 
been argued to become significant, their overall effect is to strengthen the singularity resolution effects \cite{pswe,gs1,alex}. We 
will find that the effective Hamiltonian even in the absence of these terms suffices to obtain generic resolution of strong 
singularities.} Interestingly, the polymerized Hamiltonian of LQC can also be obtained using an inverse procedure without any prior hints 
of the canonical structure or assuming a Lagrangian, just by demanding a form of repulsive nature of gravity at high curvature scales and 
general covariance \cite{ss1}.

The main results from our investigation are the following. Considering non-viscous minimally coupled matter with a general equation of 
state and anisotropic stress, we show that 
for any finite proper time, energy density is always finite in the loop quantized Kantowski-Sachs model. This is the first novel result of 
our analysis. In previous works, such as in Refs. \cite{bv,js1,jds}, energy density was found not to diverge dynamically using numerical 
simulations. Since such simulations do not cover the entire set of solutions, an analytical understanding of the behavior of energy density 
was very much needed. Our second result is to show that the physical volume remains non-zero and finite throughout the finite time evolution. Along with our result on energy density finiteness, this rules out big bang/crunch, big rip and big freeze singularities.\footnote{Unlike big bang/crunch, big rip singularity occurs at infinite volume with an infinite energy density (see e.g. \cite{ssd}). A big freeze singularity occurs at finite volume but has infinite energy density (see e.g. \cite{ps09}).}  Our third result is to show that 
even though expansion and shear scalars are universally bounded in the effective spacetime, curvature invariants can still diverge. Albeit, 
this only happens when pressure or derivatives of energy density with respect to the triads  diverge while 
energy density remains finite. Such exotic equations of state are known to cause sudden singularities in cosmological models (see e.g. 
\cite{ps09}). It turns out that the divergence in curvature invariants in finite proper time correspond to weak curvature singularities. Our fourth result is to show that all strong singularities are absent for any finite time evolution. Finally, analysis of the time-like and null geodesics 
shows that the effective spacetime is geodesically complete. There is no breakdown of geodesics for any finite proper time in the effective dynamics evolution in loop quantized Kantowski-Sachs spacetime. The effective quantum spacetime in loop quantized Kantowski-Sachs model is geodesically complete.

Our manuscript is organized in the following way. In Sec. II, we provide a brief review of the classical Hamiltonian formulation 
of the Kantowski-Sachs spacetime  in Ashtekar variables and obtain the dynamical equations. We calculate the expressions for the expansion 
and shear scalar, and for curvature invariants in terms of connection and triad variables. The effective Hamiltonian from LQC based on 
Boehmer-Vandersloot quantization is studied in Sec. III where we obtain the quantum gravitational modified  dynamical equations and obtain 
the bounded behavior of expansion and shear scalars. We obtain the analytical bounds on the triad variables from the dynamical equations for finite time evolution, 
which then imply that the energy density is finite for any finite proper time. We show that for any finite time evolution, the physical volume is non-zero and that 
the curvature invariants are non-divergent except for singular events where  pressure and energy density derivatives with respect to triads diverge at finite energy density.  In Section IV, we consider 
the special case of matter with vanishing anisotropic stress, and show that the above results turn out to be true using a simpler argument. 
Behavior of the geodesics is investigated in Sec. VA, where they are shown to not break down in any finite proper time evolution. In Sec. 
VB, 
we show that the Kantowski-Sachs spacetime in effective dynamics does not satisfy the necessary conditions for
 the existence of strong curvature singularities. Hence we conclude that the above mentioned pressure singularities are weak singularities. 
We summarize with conclusions in Sec. VI.

\section{Classical Dynamics of Kantowski-Sachs space-time}

In this section, we summarize the basic features of classical Kantowski-Sachs spacetime in Ashtekar variables. The homogeneity of the 
Kantowski-Sachs spacetime leads to a simple diagonal form for the Ashtekar-Barbero connection components and conjugate triads \cite{ab}:
\ba
A_a^i \tau_i \d x^a &=& \tilde c \tau_3 \d x + \tilde b \tau_2 \d \theta - \tilde b \tau_1 \sin\theta \d \phi + \tau_3 \cos \theta \d \phi ~,\\
\tilde E_i^a \tau_i \p_a &=& \tilde p_c \tau_3 \sin \theta \p_x + \tilde p_b \tau_2 \sin \theta \p_\theta - \tilde p_b \tau_1 \p_\phi ~,
\ea
where $\tau_i = - i \sigma_i/2$, and $\sigma_i$ are the Pauli spin matrices. The symmetry reduced triads $\tilde{p_b}$ and $\tilde{p_c}$ 
are related to the metric components of the spacetime line element:
\be
ds^2=-N^2\d t^2+g_{xx}\d x^2+g_{\Omega\Omega}\left(\d\theta^2+\sin^2{\theta}\d\phi^2\right). \label{metric}
\ee
as 
\ba
g_{xx}=\frac{\tilde{p_b}^2}{\tilde{p_c}}, ~~~~ \mathrm{and} ~~~~ g_{\Omega\Omega}=|\tilde{p_c}|. \label{gxx}
\ea
The modulus sign arises due to two orientations of the triad. Since the matter considered in this analysis is non-fermionic, we can fix one 
orientation. In the following, the orientation of the triads is chosen to be positive without any loss of generality. 
The Kantowski-Sachs spacetime is naturally foliated with spatial slices of topology $\mathbb{R} \times \mathbb{S}^2$. The spatial slices are non-compact in $x$-direction. In order to define a symplectic structure on the spatial slices, we need to restrict the integration along the $x$-direction to a fiducial length, say $L_o$. The resulting symplectic structure is:
\be
{\bf \Omega}=\frac{L_o}{2G\gamma}\left(2 \d\tilde{b}\wedge \d\tilde{p_b}+d\tilde{c}\wedge \d\tilde{p_c}\right).
\ee
where $\gamma$ is the Barbero-Immirzi parameter, its value is set to 0.2375 from black hole entropy calculations in loop quantum 
gravity. The fiducial length is a non-physical parameter in our theory, and can be arbitrarily re-scaled. In order to make the symplectic 
structure independent of  $L_o$, we introduce the new triad and connection variables $p_b$, $p_c$, and $b$, $c$ obtained by re-scaling the 
symmetry reduced triad and connection variables :
\be
p_b=L_o \tilde{p_b}, \text{  }p_c=\tilde{p_c}, \text{  }b=\tilde{b}, \text{  }c=L_o \tilde{c}. \label{scaling} ~
\ee
The non-vanishing Poisson brackets between these new variables are given by,
\ba
\left\lbrace b,p_b \right\rbrace = G \gamma, \text{            } \left\lbrace c,p_c \right\rbrace = 2G \gamma \label{poibra}.
\ea

In terms of these phase space variables, the classical Hamiltonian constraint is the following for lapse $N = 1$: 
\be
\mathcal{H}_{\rm {cl}}=\frac{-1}{2G\gamma^2}\left[2bc \sqrt{p_c}+\left(b^2+\gamma^2\right)\frac{p_b}{\sqrt{p_c}}\right]\, + \, 4\pi
p_b \sqrt{p_c} \rho \label{ham_c} ~.
\ee
Here $\rho$ is the energy density, related to matter Hamiltonian as $\rho = {\cal H}_m/V$, and $V$ is the physical volume of the fiducial 
cell: $V = 4 \pi p_b \sqrt{p_c}$. The energy density is taken to depend only on triad variables, and not on connection variables. The Hamilton's equations for the triad and connection variables are:
\begin{eqnarray}
\dot{p_b}&=&-G\gamma \frac{\partial \mathcal{H}_{\rm {cl}}}{\partial b}=\frac{1}{\gamma}\left(c\sqrt{p_c}+\frac{bp_b}{\sqrt{p_c}}\right)
\label{p_b} ,\\
\dot{p_c}&=&-2G\gamma \frac{\partial \mathcal{H}_{\rm {cl}}}{\partial c}=\frac{1}{\gamma}2b\sqrt{p_c} \label{p_c} ,\\
\dot{b}&=&G\gamma \frac{\partial \mathcal{H}_{\rm {cl}}}{\partial p_b}=\frac{-1}{2\gamma \sqrt{p_c}}\left( b^2+\gamma^2 \right) + 4 \pi 
G\gamma \sqrt{p_c} \bigg(\rho + p_b \frac{\partial \rho}{\partial p_b} \bigg) \label{clb}, \\
\dot{c}&=&2G\gamma \frac{\partial \mathcal{H}_{\rm {cl}}}{\partial p_c}=\frac{-1}{\gamma \sqrt{p_c}}\left(bc-\left( b^2+\gamma^2
\right)\frac{p_b}{2p_c} \right) + 4 \pi G\gamma \frac{p_b}{\sqrt{p_c}} \bigg(\rho + 2 p_c \frac{\partial \rho}{\partial p_c} \bigg) 
\label{clc}~.
\end{eqnarray}
Here `dot' refers to derivative with respect to proper time. 
Using the above equations, a useful result follows:
\be
\frac{d}{dt}(c p_c - b p_b)=\frac{\gamma p_b}{\sqrt{p_c}} + G \gamma V \bigg(2 p_c \frac{\partial \rho}{\partial p_c}  -p_b \frac{\partial 
\rho}{\partial p_b}\bigg) ~. \label{conserved}
\ee
As will be proved in Sec. III, it turns out that the same expression also holds in the presence of quantum gravitational modifications in 
LQC. 


Let us now find some useful expressions to understand singularities in Kantowski-Sachs spacetime. The simplest to obtain is the expression 
for energy density in terms of the gravitational phase space variables, by imposing the vanishing of the Hamiltonian constraint, 
$\mathcal{H}_{\rm{cl}} \approx 0$: 
\be
\rho = \f{1}{8 \pi G}\bigg(\f{2 b c}{\gamma^2 p_b} + \f{b^2}{\gamma^2 p_c} + \f{1}{p_c}\bigg) ~.
\ee

Two useful quantities of interest to understand the behavior of geodesics as singularities are approached are the expansion and shear 
scalars. The expansion scalar $\theta$ is given by
\be
\theta = \frac{\dot{V}}{V} = \frac{\dot{p_b}}{p_b}+\frac{\dot{p_c}}{2p_c}. \label{theta}
\ee
The shear scalar $\sigma^2$ expressed in terms of the directional Hubble rates $H_i=\dot{\sqrt{g_{ii}}}/\sqrt{g_{ii}}$ is given by
\be
\sigma^2= \frac{1}{2}\displaystyle\sum\limits_{i=1}^3 \left(H_i-\frac{1}{3}\theta\right)^2 = 
\frac{1}{3}\left(\frac{\dot{p_c}}{p_c}-\frac{\dot{p_b}}{p_b}\right)^2 . \label{sigma} ~
\ee

Next we find the expressions for curvature invariants, which when diverge signal singularities (though not necessarily strong ones). In 
terms of the reduced triad and connection variables, the expressions for the Ricci scalar, the square of the Weyl scalar and the 
Kretschmann scalar are respectively as follows:
\begin{eqnarray}
R &=&2\frac{\ddot p_b}{p_b}+\frac{\ddot p_c}{p_c}+\frac{2}{p_c} ,\label{ricciscalar} \\
C_{abcd}C^{abcd} &=&\frac{1}{3}\left[3\frac{\dot p_c}{p_c}\left(\frac{\dot p_b}{p_b}-\frac{\dot p_c}{p_c}\right)-2\left(\frac{\ddot 
p_b}{p_b}-\frac{\ddot p_c}{p_c}\right)-\frac{2}{p_c}\right]^2  \label{weylscalar} 
\end{eqnarray}
and
\begin{eqnarray}
K &=& 6\left(\frac{\dot p_b}{p_b}\frac{\dot p_c}{p_c}\right)^2-8\frac{\dot p_b}{p_b}\frac{\dot p_c}{p_c}\frac{\ddot p_b}{p_b}+4\left(\frac{\ddot p_b}{p_b}\right)^2+6\frac{\ddot p_b}{p_b}\left(\frac{\dot p_c}{p_c}\right)^2-4\frac{\ddot p_b}{p_b}\frac{\ddot p_c}{p_c} \label{kretschmannscalar} \\\nonumber
& & -8\frac{\dot p_b}{p_b}\left(\frac{\dot p_c}{p_c}\right)^3+4\frac{\dot p_b}{p_b}\frac{\dot p_c}{p_c}\frac{\ddot p_c}{p_c}+\frac{7}{2}\left(\frac{\dot p_c}{p_c}\right)^4+2\left(\frac{\dot p_c}{p_c}\right)^2\frac{1}{p_c}\\\nonumber
& & -5\left(\frac{\dot p_c}{p_c}\right)^2\frac{\ddot p_c}{p_c}+\frac{4}{p_c^2}+3\left(\frac{\ddot p_c}{p_c}\right)^2 ~.
\end{eqnarray}

We notice that the expansion and shear scalar, and all the curvature invariants depend on the following five quantities : 
$\frac{\ddot{p_b}}{p_b}$, $\frac{\ddot{p_c}}{p_c}$, $\frac{\dot{p_b}}{p_b}$, $\frac{\dot{p_c}}{p_c}$ and $\frac{1}{p_c}$. The behavior of 
$\dot p_b/p_b$ and $\dot p_c/p_c$ is obtained from the Hamilton's equations. Taking their time derivatives, we obtain
\begin{eqnarray}
\frac{\ddot{p_b}}{p_b} &=& \frac{bc}{\gamma^2 p_b} + 8 \pi G \rho + 4\pi G \bigg(p_b \frac{\partial \rho}{\partial p_b} + 2 p_c \frac{\partial \rho}{\partial p_c} \bigg) ~,\label{pbddcla} \\
\frac{\ddot{p_c}}{p_c} &=& -\frac{1}{p_c} + \frac{b^2}{\gamma^2 p_c} + 8 \pi G  \rho + 8\pi G p_b \frac{\partial \rho}{\partial p_b} 
~. \label{pcddcla}
\end{eqnarray}

It is clear from  the classical  Hamilton's equations (\ref{p_b}--\ref{clc}) and the above equations that the expansion scalar, shear 
scalar, curvature invariants and energy density all diverge as the triad components vanish, and/or the connection components diverge and/or 
the terms $\frac{\partial \rho}{\partial p_b}$ and $\frac{\partial \rho}{\partial p_c}$ diverge. 
Generic physical solutions obtained from the classical Hamiltonian constraint (\ref{ham_c}) turn out to be of 
this form and are singular.

\section{Effective loop quantum cosmological dynamics}

In the previous section, we obtained the classical singular dynamical equations from the classical Hamiltonian constraint of the 
Kantowski-Sachs spacetime. Let us now see the way quantum gravitational modifications result in non-singular dynamics. Our starting point 
is the effective Hamiltonian constraint \cite{bv}:
\be
\mathcal{H}=-\frac{p_b\sqrt{p_c}}{2G\gamma^2\Delta}\left[2\sin(b\delta_b)\sin(c\delta_c)+\sin^2(b\delta_b)+\frac{\gamma^2\Delta}{p_c}\right]+4\pi p_b\sqrt{p_c}\rho
\ee
where $\Delta$ denotes the minimum non-zero eigenvalue of the area operator in loop quantum gravity: $\Delta = 4 \sqrt{3} \pi \gamma 
\lp^2$, and 
\be
\delta_b= \sqrt{\frac{\Delta}{p_c}}, ~~~\quad ~~~\delta_c=\frac{\sqrt{\Delta p_c}}{p_b} ~.
\ee
It should be noted that the above effective Hamiltonian corresponds to the improved dynamics prescription in LQC \cite{aps3}. For 
$\delta_b$ and $\delta_c$ which are arbitrary functions of phase space variables, this prescription turns out to be unique in the sense 
that it yields physics independent of the fiducial length $L_o$ and as discussed below universal bounds on expansion and shear scalars 
\cite{js1}.

Using the Hamilton's equations, we obtain the modified dynamical equations for the gravitational phase space variables (assuming that the energy density depends only on triad variables, and not on connection variables):
\begin{eqnarray}
\dot p_b & = & \frac{p_b\cos(b\delta_b)}{\gamma \sqrt{\Delta}} \left(\sin(c\delta_c)+ \sin(b\delta_b)\right), \label{pbdot} \\
\dot p_c & = & \frac{2p_c}{\gamma \sqrt{\Delta}} \sin(b\delta_b) \cos(c\delta_c), \label{pcdot} \\
\dot b & = & -\frac{\sqrt{p_c}}{2 \gamma \Delta}\left[2\sin(b \delta_b)\sin(c \delta_c)+\sin^2(b \delta_b)+\frac{\gamma^2 
\Delta}{p_c}\right]  \nonumber\\
& & + \frac{c p_c}{\gamma \sqrt{\Delta} p_b}\sin(b \delta_b)\cos(c \delta_c) + 4 \pi G\gamma \sqrt{p_c} \bigg(\rho + p_b \frac{\partial \rho}{\partial p_b} \bigg) \label{bdot} 
\end{eqnarray}
and 
\begin{eqnarray}
\dot c & = & -\frac{p_b}{2 \gamma \Delta \sqrt{p_c}}\left[2\sin(b\delta_b)\sin(c\delta_c)+\sin^2(b\delta_b)+\frac{\gamma^2\Delta}{p_c}\right]  \nonumber\\
& & - \frac{c}{\gamma \sqrt{\Delta}}\sin(b \delta_b)\cos(c \delta_c)+\frac{b p_b}{\gamma \sqrt{\Delta} p_c} \cos(b\delta_b)\left(\sin(c\delta_c)+ \sin(b\delta_b)\right) \nonumber\\
& & +\frac{\gamma p_b}{\sqrt{p_c}} + 4 \pi G\gamma \frac{p_b}{\sqrt{p_c}} \bigg(\rho + 2 p_c \frac{\partial \rho}{\partial p_c} \bigg). 
\label{cdot}
\end{eqnarray}

As in the classical theory (see eq. \ref{conserved}), it turns out that time derivative of $(c p_c - b p_b)$ is given by 
\be
\frac{d}{dt}(c p_c - b p_b)= \nonumber \frac{\gamma p_b}{\sqrt{p_c}} + G \gamma V \bigg(2 p_c \frac{\partial \rho}{\partial p_c}  -p_b 
\frac{\partial \rho}{\partial p_b}\bigg)~. 
\ee

The change in the Hamiltonian evolution from classical theory to LQC, results in $\dot p_c$/$p_c$ and $\dot p_b$/$p_b$ as bounded 
functions. This in turn yields a non-divergent behavior of expansion and 
shear scalars. The expansion scalar is given by \cite{js1} 
\ba
\theta &=&  \frac{1}{\gamma \sqrt{\Delta}} \left(  \sin{(b\delta_b)} \cos{(c\delta_c)}+\cos{(b\delta_b)} \sin{(c\delta_c)} + 
\sin{(b\delta_b)} \cos{(b \delta_b)} \right), 
\ea
which is bounded above due to discrete quantum geometric effects inherited via area gap $\Delta$: $ |\theta| \leq 2.78/\lp$. The 
shear scalar becomes,
\ba
\sigma^2 = \frac{1}{3\gamma^2 \Delta} \left(2\sin{(b\delta_b)}\cos{(c\delta_c)}-\cos{(b\delta_b)}\left(\sin{(c\delta_c)}+\sin{(b\delta_b)} 
\right) \right)^2 ~,
\ea
which is also universally bounded \cite{js1}: $\sigma^2 \leq 5.76/\lp^{2}$. \\

From \eqref{pbdot} and \eqref{pcdot} an important result follows on the permitted values of $p_b$ and $p_c$. Let $t_0$ be some time in the 
evolution at which $p_c$ and $p_b$ have some given non-zero finite values $p_c^0$ and $p_b^0$. Then from \eqref{pcdot} we have
\begin{equation}
\int_{p_c^0}^{p_c(t)}\frac{dp_c}{p_c}= \int_{t_0}^t\frac{2}{\gamma \sqrt{\Delta}} \sin(b\delta_b) \cos(c\delta_c) \d t
\end{equation}
which implies 
\begin{equation}
p_c(t) = p_c^0  \exp\left\lbrace\frac{1}{\gamma \sqrt{\Delta}} \int_{t_0}^t\bigg(\sin(b\delta_b + c\delta_c) + \sin(b\delta_b - 
c\delta_c)\bigg) \d t\right\rbrace ~.
\end{equation}
\\
Since $ | \sin(b\delta_b + c\delta_c) + \sin(b\delta_b - c\delta_c)|\leq 2$, the integration (inside the exponential) over a finite time is 
finite. Hence, at any finite proper time, in the past or in the future : 
\begin{equation}
0<p_c(t)<\infty ~.\label{pcbound}
\end{equation}

Similarly using \eqref{pbdot}, we get
\begin{equation}
p_b(t) = p_b^0  \exp\left\lbrace\frac{1}{\gamma \sqrt{\Delta}} \int_{t_0}^t\cos(b\delta_b)\bigg(\sin(c\delta_c)+ \sin(b\delta_b)\right) 
\d t\bigg\rbrace ~.
\end{equation}
And since the integration inside the exponential is again over a bounded function, we obtain a finite integral over finite range of time. 
Hence, we obtain  
\begin{equation}
0<p_b(t)<\infty \label{pbbound}
\end{equation}
for any given finite time in past or 
future. Therefore, we reach an important result that $p_b$, $p_c$ and consequently the volume $V$ ($V=4 \pi p_b\sqrt{p_c})$ are finite, positive 
and non-zero at any finite time. Note that a similar argument was used in Ref. \cite{gowdy} to show the finiteness of the triad variables for any finite time in the effective dynamics of the Gowdy model.\\

From the vanishing of the Hamiltonian constraint, we can get the energy density in terms of dynamical variables:
\begin{equation}
\rho = \frac{1}{8\pi G\gamma^2\Delta}\left[2\sin(b\delta_b)\sin(c\delta_c)+\sin^2(b\delta_b)+\frac{\gamma^2\Delta}{p_c}\right] ~.
\end{equation}
Hence the energy density remains finite by virtue of \eqref{pcbound} and \eqref{pbbound} for any finite proper time.\\

So far we have seen that the expansion and shear scalars are generically bounded for all time.  Energy density $\rho$ remains finite under evolution over a finite proper time. And $p_b$, $p_c$ and $V$ 
remain non-zero, positive and finite under evolution over a finite proper time. The terms $\dot p_c$/$p_c$ and $\dot p_b$/$p_b$ are bounded 
due to \eqref{pcdot} and \eqref{pbdot}. Hence the divergence in curvature invariants given by \eqref{ricciscalar}, \eqref{weylscalar} and 
\eqref{kretschmannscalar} may only come from divergence in $\frac{\ddot p_c}{p_c}$ and $\frac{\ddot p_b}{p_b}$.
 After a straightforward calculation, the expressions for $\ddot p_b$ and $\ddot p_c$ turn out to be the 
following:

\begin{eqnarray}
\ddot p_b & = & p_b \Bigg[\frac{\cos(b\delta_b)\cos(c\delta_c)}{p_c} + \frac{\cos^2(b\delta_b)}{\gamma^2 \Delta}(\sin(b\delta_b)+\sin(c\delta_c))^2 \nonumber\\
& & - \frac{4 \pi}{\gamma^2 \sqrt{\Delta}}\frac{(cp_c-bp_b)}{V}\cos(c\delta_c)\bigg(\sin(c\delta_c)+\sin^3(b\delta_b)\bigg) \nonumber\\
& & +4\pi G \bigg(2p_c\frac{\partial \rho}{\partial p_c}\cos(b\delta_b)\cos(c\delta_c)-p_b\frac{\partial \rho}{\partial p_b}\sin(b\delta_b)\sin(c\delta_c) + p_b\frac{\partial \rho}{\partial p_b}\cos(2b\delta_b)\bigg)\Bigg] \label{ddotpb}
\end{eqnarray}
and
\begin{eqnarray}
\ddot p_c & = & p_c \Bigg[ -2\frac{\sin(b\delta_b)\sin(c\delta_c)}{p_c} + \frac{4\sin^2(b\delta_b)\cos^2(c\delta_c)}{\gamma^2 \Delta}\nonumber\\
& & + \frac{4 \pi}{\gamma^2 \sqrt{\Delta}}\frac{(cp_c-bp_b)}{V}\sin(2b\delta_b)\bigg(1+\sin(b\delta_b)\sin(c\delta_c)\bigg)\nonumber\\
& & + 8\pi G\bigg(p_b\frac{\partial \rho}{\partial p_b}\cos(b\delta_b)\cos(c\delta_c)-2p_c\frac{\partial \rho}{\partial p_c}\sin(b\delta_b)\sin(c\delta_c)\bigg) \Bigg] ~. \label{ddotpc}
\end{eqnarray}

The unboundedness in above terms can arise from terms containing $(cp_c-bp_b)$ and/or from terms with $\frac{\partial \rho}{\partial p_b}$ 
and $\frac{\partial \rho}{\partial p_c}$. It turns out that any potential divergences from the first type are tied to the second type in 
the following way. We have earlier found, below eq.(\ref{cdot}), that the time derivative of this difference is given by the same 
expression in the classical theory and LQC (eq.(\ref{conserved})). In 
eq.(\ref{conserved}), first term on the R.H.S is finite due to \eqref{pcbound} and \eqref{pbbound}. Integrating the right hand side, we see 
that the quantity $(cp_c-bp_b)$ may diverge if the derivatives of the energy density with respect to triads diverge. Otherwise $(cp_c-bp_b)$ 
will be finite at any finite past or future time.

In conclusion, any divergences in $\frac{\ddot p_c}{p_c} $ and $\frac{\ddot p_b}{p_b}$, and consequently in the curvature 
invariants come from the terms $\frac{\partial \rho}{\partial p_b}$ and $\frac{\partial \rho}{\partial p_c}$. Since energy density is 
always finite for any finite time, we need matter with an equation of state which has divergent triad derivatives of 
energy density with energy density being finite. Only then $\ddot p_b$ and $\ddot p_c$ can diverge in finite time in this loop quantized 
Kantowski-Sachs model. And only then the curvature invariants can diverge. These divergences in curvature invariants in case of 
special matter types lead to weak curvature singularities as will be shown in section V. For all other types of matter the curvature 
invariants are non-divergent for all finite times indicating the absence of any singularities.

\section{Effective Dynamics : matter with vanishing anisotropic stress and pressure singularities}
In the previous section we found that for effective spacetime description in LQC, the only way curvature invariants can diverge is when derivatives of energy density with respect to triads diverge at finite energy density. We will now show that for the special case of matter having a vanishing anisotropic stress, these divergences are related to the pressure divergences. For such a matter, the energy density is a function of volume only, i.e. $\rho(p_b,p_c)=\rho(p_b\sqrt{p_c})$. Then the pressure $P$ can be written as,
\be
P= -\frac{\partial \mathcal{H}_{\rm {matt}}}{\partial V} = -\rho - V\frac{\partial \rho}{\partial V} ~.
\ee
The derivatives of the energy density with respect to triads can be written in terms of the energy density and pressure as:
\begin{equation}
p_b\frac{\partial \rho}{\partial p_b}=2p_c\frac{\partial \rho}{\partial p_c}=-\rho-P , \label{pbder}
\end{equation}

Using the above equations, the expression for $\ddot p_b$ in the case of vanishing anisotropic stress can be obtained from eq.(\ref{ddotpb}). It turns out to be:
\begin{eqnarray}
\ddot p_b & = & p_b \Bigg[\frac{\cos(b\delta_b)\cos(c\delta_c)}{p_c} + \frac{\cos^2(b\delta_b)}{\gamma^2 \Delta}(\sin(b\delta_b)+\sin(c\delta_c))^2 \nonumber\\
& & - \frac{4 \pi}{\gamma^2 \sqrt{\Delta}}\frac{(cp_c-bp_b)}{V}\cos(c\delta_c)\bigg(\sin(c\delta_c)+\sin^3(b\delta_b)\bigg) \nonumber\\
& & -4\pi G \bigg(\cos(b\delta_b+c\delta_c)+\cos(2b\delta_b)\bigg)(\rho+P)\Bigg] ~.
\end{eqnarray}
Similarly, eq.(\ref{ddotpc}) yields,
\begin{eqnarray}
\ddot p_c & = & p_c \Bigg[ -2\frac{\sin(b\delta_b)\sin(c\delta_c)}{p_c} + \frac{4\sin^2(b\delta_b)\cos^2(c\delta_c)}{\gamma^2 \Delta}\nonumber\\
& & + \frac{4 \pi}{\gamma^2 \sqrt{\Delta}}\frac{(cp_c-bp_c)}{V}\sin(2b\delta_b)\bigg(1+\sin(b\delta_b)\sin(c\delta_c)\bigg)\nonumber\\
& & - 8\pi G\bigg(\cos(b\delta_b+c\delta_c)\bigg)(\rho+P)\Bigg] ~.
\end{eqnarray}

Before we analyze the nature of potential singularities, it is interesting to note that the time derivative of $(c p_c - b p_b)$ in case of matter with vanishing anisotropic stress is given by 
\be
\frac{d}{dt}(cp_c-bp_b)=\frac{\gamma p_b}{\sqrt{p_c}} ~.\label{cpcbpb1}
\ee
This is easily checked by using eq.(\ref{pbder}) in eq.(\ref{conserved}). The right hand side of \eqref{cpcbpb1} is bounded by virtue of \eqref{pcbound} and \eqref{pbbound}, which implies that the quantity $(cp_c-bp_b)$ is also bounded at any finite past or future time.\\

We have shown earlier in Sec. III that $p_b$, $p_c$ and $V$ remain non-zero, positive and finite under evolution over a finite proper time. Equation \eqref{cpcbpb1} implies that $(cp_c-bp_c)$ is finite upon evolution over a finite time. Hence in effective dynamics in LQC for matter with a vanishing anisotropic stress, both $\frac{\ddot p_c}{p_c}$ and $\frac{\ddot p_b}{p_b}$, and consequently the curvature invariants given by \eqref{ricciscalar}, \eqref{weylscalar} and \eqref{kretschmannscalar} are non-divergent except when the pressure diverges at a finite value of energy density, shear scalar and expansion scalar and non-zero volume. In the next section, we would show that such pressure singularities are weak singularities and geodesics evolution does not break down at such events.

\section{Analysis of geodesics and strength of possible singularities}
In this section, we analyze whether the effective spacetime description of the Kantowski-Sachs model in LQC results in geodesic evolution which 
breaks down in finite time, and the strength of potential singularities. We start with an analysis of geodesics, followed by the strength of singularities using Kr\'{o}lak's condition \cite{krolak}. 

\subsection{Geodesics}

We noted in the previous section that the curvature invariants are generically bounded in effective dynamics except for very specific type of pressure singularities. This means that there may be potential singularities in the effective spacetime description of Kantowski-Sachs spacetime. A commonly used criterion to characterize singularities is that all the geodesics that go into the singularity must end at the singularity, i.e. the geodesics must not be extendible beyond the singularity. However, if geodesics can be extended beyond the point where the curvature invariants diverge, then it may not be a strong enough singularity to be physically significant.\footnote{For examples in GR and LQC, see Refs. \cite{jambrina} and \cite{ps09} respectively.} The spacetime may be extendable in such a case. Geodesic (in)completeness analysis is therefore important to 
understand the exact nature of singularities or lack thereof. 


For the metric of the Kantowski-Sachs spacetime \eqref{metric}, the geodesic equations yield the following second order equations in the affine parameter $\tau$:
\begin{equation}
\left(g_{xx}x'\right)'= 0, ~~~~~~ \left(g_{\Omega\Omega}\sin^2(\theta)\phi'\right)' = 0,
\end{equation}
\begin{equation}
\left(g_{\Omega\Omega}\theta'\right)' = g_{\Omega\Omega}\sin\theta\cos\theta \phi'^2,
\end{equation}
and
\begin{equation}
-2t't''=g_{xx}'x'^2+g_{\Omega\Omega}'(\theta'^2 + \sin^2\theta \phi'^2) ~.
\end{equation}
Here prime denotes derivative with respect to the affine parameter. And, we 
recall that the metric components, $g_{xx}$ and $g_{\Omega\Omega}$ are related to the triads $p_b$ and $p_c$ via eq.(\ref{gxx}).

To find the solutions, we notice that one can rotate angular coordinates in such a way that initially when affine parameter $\tau=0$, $\theta(0)=\pi /2$ and $\theta'(0)=0$. Then  $\theta(\tau)=\pi /2$ for all $\tau$ is a solution of the above $\theta$ geodesic equation with these initial conditions. Due to the uniqueness of solutions of second order differential equations with given initial conditions, this is the unique solution. Therefore, we will assume that $\theta=\pi /2$ hereafter. Using this result, the solutions to the remaining geodesic equations in $x$, $\phi$ and $t$ are:
\begin{equation}
x'=\frac{C_x}{g_{xx}}, ~~~~~ \phi'=\frac{C_{\phi}}{g_{\Omega\Omega}} ~, \label{geodesics}
\end{equation}
and
\begin{equation}
t'^2=\epsilon + \frac{C_x^2}{g_{xx}}+ \frac{C_{\phi}^2}{g_{\Omega\Omega}} ~. \label{tprime}
\end{equation}
Here $C_x$ and $C_{\phi}$ are constants of integration, and  $\epsilon=1$ for timelike geodesics and $\epsilon=0$ for null geodesics.

In classical GR, the geodesic equations break down if either $g_{xx}$ or $g_{\Omega\Omega}$ vanishes at a finite value of the affine parameter. This is certainly the case for the classical singularity in the Kantowski-Sachs spacetime which results in geodesic incompleteness. The situation changes dramatically, when quantum gravitational effects in LQC are in play. Due to the bounds on the values of $p_b$ and $p_c$ given in \eqref{pbbound} and \eqref{pcbound}, both $g_{xx}$ and $g_{\Omega\Omega}$, as defined in equation \eqref{gxx}, are finite, non-zero, positive functions for any finite time. This implies that the geodesic evolution never breaks down in effective dynamics in loop quantized Kantowski-Sachs model. For any finite time evolution, effective spacetime is geodesically complete.

\subsection{Strength of Singularities}

Apart from the analysis of geodesics, important information about the nature of singularities can be found by analyzing their strength. This is determined  by considering what happens to an object as it falls into the singularity. A strong curvature singularity is defined  as one that crushes any in-falling objects to zero volume irrespective of the properties or composition of the objects \cite{Ellis,tipler}. Basically the curvature squeezes any in-falling objects to infinite density. Infinite tidal forces completely destroy any arbitrary in-falling object. In contrast to the strong singularities, weak singularities do not imply a complete destruction of the in-falling objects. Even though some curvature components or curvature invariants may diverge, it is possible to construct a sufficiently strong detector which survives large tidal forces and escapes the singular event. These qualitative notions has been put in precise mathematical terms by Tipler \cite{tipler} and Kr\'{o}lak \cite{krolak}. It has been 
conjectured that the physical singularities in the sense of geodesic incompleteness are those which are also strong curvature type \cite{tipler,krolak}. It has been argued that if the conjecture is satisfied then a weak form of Penrose's cosmic  censorship hypothesis can be proved \cite{krolak}.

The necessary conditions for a strong curvature singularity derived by Kr\'{o}lak are broader than Tipler's conditions. Any singularity which is weak by Kr\'{o}lak's conditions will be weak by Tipler criteria, but the converse is not true. So we use the Kr\'{o}lak conditions in order to search for the signs of strong singularities in the broadest sense. According to necessary conditions due to Kr\'{o}lak \cite{krolak}, if a singularity is a strong curvature singularity, then for some non-spacelike geodesic running into the singularity, the following integral diverges as the singularity is approached:

\begin{equation}
K^{i}_{j} =\int_0^{\tau} d\tilde\tau |R^{i}_{4j4} (\tilde\tau)| ~. \label{krolak1}
\end{equation}

That is, if there is a strong curvature singularity in the region, then for a non-spacelike geodesic running into the singularity the following necessary condition is satisfied:
\begin{equation}
\lim_{\tau \to\tau_o}K^{i}_{j} \rightarrow \infty  \label{krolak}
\end{equation}
where $\tau_o$ is the value of the affine parameter at which the singularity is located.\\

Considering the behavior of the integrand, i.e. components of Riemann tensor, can lead us to understand which terms may potentially diverge and result in strong singularities. The non-zero components of the Riemann tensor for the Kantowski-Sachs metric in terms of the triads are:

\begin{eqnarray}
R^1_{212}&=&g_{xx}R^2_{121} \nonumber \\
&=&\bigg(\frac{p_b^2}{L_o^2 p_c} \bigg) \bigg[ -\frac{3}{4} \bigg(\frac{\dot p_c}{p_c} \bigg)^2 - \bigg(\frac{\ddot p_b}{p_b} \bigg)
+ \bigg(\frac{\dot p_b}{p_b} \bigg)\bigg(\frac{\dot p_c}{p_c} \bigg) + \frac{1}{2}\bigg(\frac{\ddot p_c}{p_c} \bigg) \bigg] ,\\
R^1_{441}&=& \sin^2\theta R^1_{331} = p_c\sin^2\theta R^3_{131} = p_c\sin^2\theta R^4_{141} \nonumber \\
&=& p_c\sin^2\theta \bigg[ \frac{1}{4}\bigg(\frac{\dot p_c}{p_c} \bigg)^2 - \frac{1}{2}\bigg(\frac{\ddot p_c}{p_c} \bigg) \bigg], \\
R^2_{442}&=&\sin^2\theta R^2_{332}=-\frac{p_c}{g_{xx}}R^3_{232}=-\frac{p_c}{g_{xx}}R^4_{242}\nonumber \\
&=&-p_c \sin^2\theta \bigg[\frac{1}{2}\bigg(\frac{\dot p_b}{p_b} \bigg)\bigg(\frac{\dot p_c}{p_c} \bigg) - \frac{1}{4}\bigg(\frac{\dot 
p_c}{p_c} \bigg)^2 \bigg], 
\end{eqnarray}
and,
\begin{equation}
R^3_{443}=-\sin^2\theta R^4_{343} 
= -\sin^2\theta \bigg[ 1+\frac{p_c}{4}\bigg(\frac{\dot p_c}{p_c} \bigg)^2  \bigg] ~.
\end{equation}
Note that the factors of $\sin^2\theta$ can be ignored in this analysis, as we  
 can always choose $\theta=\pi /2$ along our geodesics as discussed in Sec. VA.

Most of the terms in all the Riemann tensor components are of the type $\bigg(\frac{\dot p_1}{p_1}\bigg)^m \bigg(\frac{\dot p_2}{p_2}\bigg)^n \bigg(\frac{\dot p_3}{p_3}\bigg)^q f(p_1,p_2,p_3)$, which are made out of products of powers of $p_b, p_c, \frac{\dot p_c}{p_c}$ and $\frac{\dot p_b}{p_b}$, which are functions of the affine parameter. The other type of terms are $g(p_b,p_c)\frac{\ddot p_c}{p_c}$ or $g(p_b,p_c)\frac{\ddot p_b}{p_b}$, where $g(p_b,p_c)$ is a function only of $p_b$ and $p_c$ without involving any of their derivatives.

First note that the integral in \eqref{krolak1} involves an integral of the absolute value of Riemann tensor components, and in turn each Riemann tensor component is a sum of several terms. Since the integral of the absolute value of a sum is always less than or equal to the integral of the sum of the absolute value of each term, for our purposes it would suffice to look individually at the integrals of the absolute value of each term separately. So we consider the different types of terms present in the Riemann tensor components mentioned above one by one.

We first look at terms of type $\bigg(\frac{\dot p_1}{p_1}\bigg)^m \bigg(\frac{\dot p_2}{p_2}\bigg)^n \bigg(\frac{\dot p_3}{p_3}\bigg)^q f(p_1,p_2,p_3)$. We can split the integral from $0$ to $\tau_o$ into pieces where the integrand takes a definite sign (positive or negative), e.g.

\begin{eqnarray}
\int_0^{\tau_o} d\tau \bigg| \bigg(\frac{\dot p_1}{p_1}\bigg)^m \bigg(\frac{\dot p_2}{p_2}\bigg)^n \bigg(\frac{\dot p_3}{p_3}\bigg)^q f(p_1,p_2,p_3)\bigg| &=& \int_0^{\tau_1} d\tau \bigg(\frac{\dot p_1}{p_1}\bigg)^m \bigg(\frac{\dot p_2}{p_2}\bigg)^n \bigg(\frac{\dot p_3}{p_3}\bigg)^q f(p_1,p_2,p_3) \nonumber \\
& & - \int_{\tau_1}^{\tau_2} d\tau \bigg(\frac{\dot p_1}{p_1}\bigg)^m \bigg(\frac{\dot p_2}{p_2}\bigg)^n \bigg(\frac{\dot p_3}{p_3}\bigg)^q f(p_1,p_2,p_3) \nonumber \\
& & + \int_{\tau_2}^{\tau_3} d\tau \bigg(\frac{\dot p_1}{p_1}\bigg)^m \bigg(\frac{\dot p_2}{p_2}\bigg)^n \bigg(\frac{\dot p_3}{p_3}\bigg)^q f(p_1,p_2,p_3) \nonumber \\
& & . \nonumber \\
& & . \nonumber \\
& & . \nonumber \\
& & + \int_{\tau_n}^{\tau_0} d\tau \bigg(\frac{\dot p_1}{p_1}\bigg)^m \bigg(\frac{\dot p_2}{p_2}\bigg)^n \bigg(\frac{\dot p_3}{p_3}\bigg)^q f(p_1,p_2,p_3) \nonumber
\end{eqnarray}
Now focus on any one of the terms from the above expression, say $\tau_k$ to $\tau_{k+1}$,
\begin{eqnarray}
\int_{\tau_k}^{\tau_{k+1}} d\tau \bigg(\frac{\dot p_1}{p_1}\bigg)^m \bigg(\frac{\dot p_2}{p_2}\bigg)^n \bigg(\frac{\dot p_3}{p_3}\bigg)^q f(p_1,p_2,p_3)&=&\int_{t_k}^{t_{k+1}}  dt \frac{d\tau}{dt} \bigg(\frac{\dot p_1}{p_1}\bigg)^m \bigg(\frac{\dot p_2}{p_2}\bigg)^n \bigg(\frac{\dot p_3}{p_3}\bigg)^q f(p_1,p_2,p_3) \nonumber \\[10pt] 
&=&\int_{t_k}^{t_{k+1}}  dt \frac{\bigg(\frac{\dot p_1}{p_1}\bigg)^m \bigg(\frac{\dot p_2}{p_2}\bigg)^n \bigg(\frac{\dot p_3}{p_3}\bigg)^q f(p_1,p_2,p_3)}{\sqrt{\epsilon + \frac{C_x^2 L_o^2 p_c}{p_b^2} + \frac{C_{\phi}^2}{p_c}}} ~.\label{fp1}
\end{eqnarray}
Here we have used eq. \eqref{tprime}, and note that $\epsilon$ is $1$ for timelike geodesics and $0$ for null geodesics.  We have shown earlier in Sec. III, particularly equations \eqref{pcdot}, \eqref{pbdot}, \eqref{pcbound} and \eqref{pbbound} that $\frac{\dot p_b}{p_b}$ and $\frac{\dot p_c}{p_c}$ are bounded functions, and $p_b, p_c$ are non-zero and finite as well for all finite values of the coordinate time $t$. The quantity under the square root in the denominator of equation \eqref{fp1} is therefore positive definite because of the bounds on $p_b$ and $p_c$. In the following, let us consider the special case when both of the integration constants of the geodesic equations, $C_x$ and $C_{\phi}$, happen to be simultaneously zero.

In the timelike case, since $\epsilon$ is equal to unity, we see from \eqref{geodesics} and \eqref{tprime} that it represents the worldline of a massive particle sitting at rest at a location in space, and the denominator in \eqref{fp1} becomes unity. However, in the case of null geodesics (photons), since $\epsilon$ is zero it seems that the 
denominator in the R.H.S. of \eqref{fp1} could be zero if both $C_x$ and $C_{\phi}$ vanish simultaneously. But we find from \eqref{geodesics} and \eqref{tprime} that if both $C_x$ and $C_{\phi}$ are simultaneously zero, then we have the peculiar situation with coordinates $x,\phi$ and $t$ being constant as a function of the affine parameter. This means that the whole geodesic will be just one event in the spacetime manifold, i.e. the photon appears for one moment at some location and disappears immediately. Such a case is hence not relevant for our discussion of the strength of singularities as it does not correspond to a physically suitable null geodesic. 

Thus, we show that the integrand in R.H.S. of \eqref{fp1} is well-defined, real and finite for all finite values of the time $t$. That means that the integral will be finite if the upper limit $t_o$ is finite. 


If initially both $\tau$ and $t$ start from zero, then
\begin{equation}
\tau_0=\int_0^{\tau_o}d\tau=\int_0^{t_o} dt \frac{d\tau}{dt}=\int_0^{t_o} \frac{dt}{\sqrt{\epsilon + \frac{C_x^2 L_o^2 p_c}{p_b^2} + 
\frac{C_{\phi}^2}{p_c}}} .\label{tau_o}
\end{equation}
Note that for observers comoving with respect to the matter world lines (the fundamental observers), the proper time is given by $t$, hence  
 $\tau_0 = t_0$. Hence for a finite $\tau_0$, $t_0$ is always finite for such observers. The integral in \eqref{fp1} is then finite for finite $\tau_0$. In general, the integrand on the R.H.S. of \eqref{tau_o} is positive definite and finite for finite $t$. It is possible that for certain geodesics there can be potential cases where the upper limit $t_0$ is infinite even when $\tau_0$ is finite. If such a case exists and if the integral \eqref{fp1} diverges in such a case, then this divergence occurs at an infinite proper time for fundamental observers. Further, for such a potential divergence the energy density is still finite in the finite time evolution for the matter world lines (using the results from Sec. III). Hence, such a potential divergence would not correspond to any known strong singularity such as big bang/crunch, big rip and big freeze singularities which are characterized by divergence in energy density in finite proper time for fundamental observers.
 Thus, we can conclude that terms in the Riemann curvature components of the type $\bigg(\frac{\dot p_1}{p_1}\bigg)^m \bigg(\frac{\dot p_2}{p_2}\bigg)^n \bigg(\frac{\dot p_3}{p_3}\bigg)^q f(p_1,p_2,p_3)$ in \eqref{krolak} will not contribute to any potential divergences in finite proper time measured by comoving observers. 

Let us now consider the other type of terms in the Riemann tensor components. These are of the type $g(p_b,p_c)\frac{\ddot p_c}{p_c}$ or $g(p_b,p_c)\frac{\ddot p_b}{p_b}$ with $g(p_b,p_c)$ independent of any derivatives. Again as before, we split the integral of the absolute value into pieces where the integrand has a definite sign, and then look at one of those pieces. These terms will be integrated at least once in \eqref{krolak}, and it can be seen that on integrating by parts, there are no terms left with double derivatives of $p_b$ or $p_c$. For example,

\begin{eqnarray}
\int g(p_b,p_c)\frac{\ddot p_c}{p_c}d\tau  &=& \int \frac{g(p_b,p_c)}{\sqrt{\epsilon + \frac{C_x^2 L_o^2 p_c}{p_b^2} + \frac{C_{\phi}^2}{p_c}}}\frac{\ddot p_c}{p_c}dt =: \int g_1(p_b,p_c)\ddot p_c dt \nonumber \\
&=& g_1(p_b,p_c)\dot p_c - \int \dot p_c \bigg(\frac{d g_1(p_b,p_c)}{dt}\bigg)dt \nonumber \\
&=& g_2(p_b, p_c)\frac{\dot p_c}{p_c} - \int f_2(p_b, p_c, \frac{\dot p_c}{p_c},\frac{\dot p_b}{p_b}) dt .\label{div}
\end{eqnarray} 
Here in the last line, we have defined $g_2(p_b, p_c) := p_c g_1(p_b,p_c)$, and $f_2(p_b, p_c, \frac{\dot p_c}{p_c},\frac{\dot p_b}{p_b}) := \dot p_c \dot g_1(p_b,p_c)$. Note that $f_2(p_b, p_c, \frac{\dot p_c}{p_c},\frac{\dot p_b}{p_b})$ is a term of type $\bigg(\frac{\dot p_1}{p_1}\bigg)^m \bigg(\frac{\dot p_2}{p_2}\bigg)^n \bigg(\frac{\dot p_3}{p_3}\bigg)^q f(p_1,p_2,p_3)$.
We have already shown in \eqref{fp1} that integrals of terms like $f_2(p_b, p_c, \frac{\dot p_c}{p_c},\frac{\dot p_b}{p_b})$ over a finite range of proper time for fundamental observers are non-divergent. And $g_2(p_b, p_c)\frac{\dot p_c}{p_c}$ is finite for finite values of time $t$. As for the case of \eqref{fp1}, integral \eqref{div} does not result in a divergence occuring in a finite proper time for comoving observers. The terms containing quantities like $\frac{\ddot p_c}{p_c}$ or $\frac{\ddot p_b}{p_b}$, which could lead to potential divergences arising from pressure or derivative of energy density with respect to the triads, as mentioned in Secs. III and IV, are removed upon integrating once.

Hence we have established that the necessary condition \eqref{krolak} for the presence of strong curvature singularity, i.e equation \eqref{krolak}, is not satisfied in effective dynamics of Kantowski-Sachs spacetime for any finite proper time measured by fundamental observers. Therefore, all the potential curvature divergent events associated with pressure and derivatives of energy density with triads  discussed in Secs. III and IV turn out to be weak singularities. 

\section{Conclusions}

A key question for any quantum theory of gravity is whether it can successfully resolve various spacelike singularities. As classical singularities are 
generic features of the classical continuum spacetime, the analogous question is whether non-existence of singularities is a generic result of quantum spacetime. As there are singularity theorems in  classical GR, is there an analogous non-singularity theorem in quantum gravity? Since we do not have a 
full theory of quantum gravity, these questions can not be fully answered at the present stage. Yet, valuable insights can be gained by understanding whether and how quantum gravitational effects lead to singularity resolution in spacetimes which can be quantized. By systematically studying such spacetimes with increasing complexity, one expects that key features of singularity resolution in general in quantum gravity can be uncovered. 

Loop quantum cosmology provides a very useful avenue for these studies. In recent years, a rigorous quantization of various spacetimes has 
been performed, and resolution of singularities in different models has been found \cite{as}. The effective spacetime description of LQC 
enables us to understand singularity resolution in considerable detail. In previous works, using this description above questions on generic 
resolution of singularities have been addressed in isotropic and Bianchi-I spacetimes \cite{ps09,ps11,ps15}. In these works, it was found 
that in the effective spacetime description of LQC all strong singularities are resolved and spacetime is geodesically complete. Our goal 
in this manuscript was to probe these issues in Kantowski-Sachs spacetime in LQC using Boehmer-Vandersloot prescription \cite{bv}. 

In contrast to the previous investigations on this topic, Kantowski-Sachs spacetime is additionally non-trivial. Unlike the isotropic and Bianchi-I spacetime in LQC, energy density is not universally bounded because of the presence of inverse power of a triad component $(p_c)$. Note that universal bound on energy density played an important role in proving geodesic completeness and resolution of strong singularities in isotropic and Bianchi-I spacetimes \cite{ps09,ps11,ps15}. It was recently found using numerical simulations that dynamical bounds exist on energy density in loop quantized Kantowski-Sachs model \cite{js1,js2}. However, an analytical proof was needed to reach general conclusions about singularity resolution. A novel result in our analysis is that in any finite time range, energy density remains finite. Coupled with another result from our present analysis, that volume never becomes zero or infinite in finite time evolution, we find that singularities such as big bang/crunch which occur at zero 
volume with infinite energy density, big rip singularities occurring at infinite volume with infinite energy density, and big freeze singularities occurring at finite volume but infinite energy density are avoided.

The finiteness of energy density, expansion and shear scalars does not imply that curvature invariants are also finite. Investigating their behavior, we find that the curvature invariants remain bounded for all finite times except under certain circumstances. If the matter present is such that the derivatives of the energy density with respect to the triad variables can diverge even though the energy density is finite, then the curvature invariants diverge at these events. By considering the case of matter with vanishing anisotropic stress we show that these triad-derivatives of energy density are related to the pressure. In other words these divergences occur due to divergences in pressure at finite value of energy density. Do these events where curvature invariants diverge 
imply strong singularities and geodesic incompleteness? The answer turns out to be negative.

Analyzing geodesics to understand the nature of the potential singularities indicated by divergences in curvature invariants, we find that 
the Kantowski-Sachs spacetime is geodesically complete in the effective dynamics of LQC. That means geodesics can be extended beyond the 
potential singularities where pressure or triad derivatives of energy density diverge at finite energy density, expansion and shear scalars. 
Using Kr\'{o}lak's condition of the strength of the singularities, these potential singularities turn out to be weak. We find that all 
known strong curvature singularities are non-existent in finite time evolution in effective spacetime.
Thus, the only possible singularities in effective spacetime of Kantowski-Sachs model in LQC are weak singularities beyond which geodesic 
can be extended.  

Our analysis, thus generalizes previous results on geodesic completeness and strong singularity resolution in LQC to Kantowski-Sachs spacetime providing 
useful insights on singularity resolution in black hole interior and in presence of anisotropies and spatial curvature. Note that our 
analytical results though show strong singularity avoidance in any finite time evolution, the question of 
how exactly the singularity is resolved for a specific matter can be answered only using numerical simulations. Such numerical 
investigations carried out for scalar fields, massless and in presence of potentials, show that classical singularity is replaced by  
bounces of triads \cite{bv,chiou,js1,js2}. All these results obtained in different spacetimes imply robust signs of 
quantum geometric effects as understood in loop quantum gravity yielding a generic resolution of strong singularities. Future 
investigations with more complex and richer spacetimes are important in this direction.

\begin{acknowledgments}
We thank an anonymous referee for useful comments and suggestions on the manuscript which led to its improvement. This work is supported by NSF grants PHYS1404240 and PHYS1454832.
\end{acknowledgments}

\end{document}